# Polymer Derived SiOC Integrated with Graphene Aerogel as Highly Stable Li-ion Battery Anodes

*Gaofeng Shao[a,b], Dorian A. H. Hanaor [b,*], Jun Wang[b], Delf Kober[b], Shuang Li[c], Xifan Wang[b], Xiaodong Shen[d], Maged F. Bekheet[b], Aleksander Gurlo[b]*

a.  Institute of Advanced Materials and Flexible Electronics (IAMFE), School of Chemistry and Materials Science, Nanjing University of Information Science & Technology, 210044, Nanjing, China
b.  Fachgebiet Keramische Werkstoffe / Chair of Advanced Ceramic Materials, Technische Universität Berlin, 10623, Berlin, Germany
c.  Functional Materials, Department of Chemistry, Technische Universität Berlin, Berlin 10623, Germany
d.  College of Materials Science and Engineering, Nanjing Tech University, 211816, Nanjing, China

*Corresponding author, E–mail dorian.hanaor@ceramics.tu-berlin.de

https://doi.org/10.1021/acsami.0c12376

**Abstract**: Amorphous polymer-derived silicon oxycarbide (SiOC) is an attractive candidate for Li-ion battery anodes, as an alternative to graphite, which is limited to a theoretical capacity of 372 mAh/g. However, SiOC tends to exhibit poor transport properties and cycling performance as the result of sparsely distributed carbon clusters and inefficient active sites. To overcome these limitations, we designed and fabricated a layered graphene/SiOC heterostructure by solvent assisted infiltration of a polymeric precursor into a modified 3D graphene aerogel skeleton. The use of a high-melting-point solvent facilitated the precursor's freeze-drying, which following pyrolysis yielded SiOC as a layer supported on the surface of nitrogen doped reduced graphene oxide aerogels. The fabrication method employed here modifies the composition and microstructure of the SiOC phase. Among the studied materials, highest levels of performance were obtained for a sample of moderate SiOC content, in which the graphene network constituted 19.8wt% of the system. In these materials a stable reversible charge capacity of 751 mAh/g was achieved at low charge rates. At high charge rates of 1480 mA/g the capacity retention was ~95% (352 mAh/g) after 1000 consecutive cycles. At all rates, Colombic efficiencies >99% were maintained following the first cycle. Performance across all indicators was majorly improved in the graphene aerogel/SiOC nanocomposites, compared with unsupported SiOC. Performance was attributed to mechanisms across multiple length-scales. The presence of oxygen rich $SiO_{4-x}C_x$ tetrahedra units and a continuous free-carbon network within the SiOC provide sites for reversible lithiation, while high ionic and electronic transport is provided by the layered graphene/SiOC heterostructure.

**1. Introduction**

The development of high-performance anode materials for use in Li-ion batteries has driven significant research into carbon and silicon based nanostructures[1-7]. However, the achievement of increasingly higher reversible capacities, improved cycling behaviour and rapid charging remains a challenging objective. The material most commonly used for Li-ion battery anodes is graphite. However, graphite electrodes are limited to a theoretical capacity of 372 mAh/g for fully lithiated graphite ($LiC_6$)[7-8]. In recent years, higher capacities have been achieved in carbon nanotubes and graphene based systems, however purely carbon based anodes are not conducive to large reversible capacities at high current densities[9]. Silicon has attracted attention owing to its significantly higher theoretical maximum capacity of 3579 or 4200 mAh/g (lithiated to $Li_{15}Si_4$ or $Li_{4.4}Si$ respectively) [7, 10-12]. However, performance in silicon materials is often hindered by high volumetric expansion during lithiation as well as poor ionic mobility[12]. The integration of polymer derived ceramics (PDCs) in anode materials for lithium ion batteries is an attractive prospect that has been the focus of numerous studies since early work into lithiation of PDCs, reported in 1994[13]. PDCs, which are generally carbide and oxycarbide materials formed by the pyrolysis of silicon based preceramic polymers[14-17], are of growing interest towards diverse applications including electrochemical energy storage[18-23], microbial bioelectrochemical systems[24], microelectromechanical systems[25], electromagnetic shielding[26] and thermoelectrics[27]. The development of PDCs is motivated by their remarkable chemical stability and the versatile routes with which they can be processed[28-29]. Coupled with tuneable properties of electrical conductivity, permittivity and thermal conductivity, PDCs offer utilitarian pathways towards effective energy materials[16-17, 30].

Silicon oxycarbide, referred to as SiOC or oxycarbide glass, is an amorphous ceramic that is readily formed from preceramic polymers with varying stoichiometry. At an atomic scale SiOC is comprised of short range ordered $SiO_{4-x}C_x$ tetrahedra alongside a phase of free carbon. However, two notable obstacles remain towards the application of SiOC and other PDCs in Li-ion battery anodes[31]; (1) A low intrinsic electronic conductivity mediated by disparate nanosized free carbon clusters impedes Li-ion insertion/deinsertion reactions in PDCs [32]. (2) Dense PDC microstructures that result from most processes lack ion transport pathways and generally yield poor utilisation of the active phase. To address these shortcomings, the composition and structure of applied PDCs can be designed at various scales to provide improved electronic and ionic transport behaviour in electrodes. PDCs have been processed in forms of etched porous particles[33], aerogels[34], beads[21, 35], and fibers[36-37] to impart improved structure driven performance. In terms of composition, PDCs with increased free carbon have been designed through control of pyrolysis conditions[38] and the use of carbon-rich precursors or cross-linkers[39-40]. Composite materials incorporating carbon nanostructures (CNTs[41] or graphene[37, 42-44]) and /or nanoparticles of Sn[18, 23], Sb[19, 45], or Si[10, 20] into SiOC offer further pathways towards improved electrochemical performance.

Enhanced electrical conductivity and electrochemical performance have been observed in mixtures of graphene or graphene oxide with PDCs formed either by physical mixing or chemical synthesis, with subsequent densification[26, 46-47]. However, uniformly dispersing graphene in PDCs without aggregation is challenging and thus flawed microstructures are generally formed. In graphene/PDC composite anodes, the improvement of electrical properties is often limited as the result of low graphene loading, its disordered distribution and the absence of conductive pathways[42, 46, 48].

Monolithic 3D graphene-based structures developed in recent years open new opportunities for the fabrication of dense graphene/ceramic composites with novel properties[49-51]. In the study reported here, highly porous, electrically conductive, 3D structured graphene aerogels were embedded with polymer-derived SiOC. A high-melting-point-solvent assisted infiltration of a preceramic polymer precursor into the interconnected 3D scaffold of a graphene aerogel was followed by freeze-drying and high temperature pyrolysis to yield a new form of graphene/PDC composite. At multiple length-scales, the structures of materials thus obtained provide pathways towards highly effective transport and lithiation performance.

In the materials reported here, a robust reduced graphene oxide aerogel acts as 3D space-confined matrix, producing 2D layered graphene/SiOC heterostructures at the microscale, while further modifying the structure of $SiO_{4-x}C_x$ tetrahedral units by residual oxygen groups, thus endowing the SiOC component with continuous carbon nano-domains and conferring improved electronic transport in the PDC. At higher length-scales, in the bulk graphene/SiOC composite, an interconnected network structure is formed, in which thin SiOC layers are homogeneously coupled with multilayer graphene sheets. Overall, this synthesis approach leads to materials with favourable levels of electrical conductivity and electrochemical performance in Li-ion batteries.

## 2. Experimental

### 2.1 Fabrication of 3D graphene-based aerogels

To produce composite materials, we first synthesised a highly porous skeleton of N-doped reduced graphene oxide aerogel (NRGOA) via a polymerization-reduction process with subsequent freeze drying and thermal reduction. In a typical process, graphene oxide (GO) was prepared from natural graphite by an improved Hummers method[52] and ultrasonically dispersed (5 mg/ml) in cold deionized water for 2 h. An aqueous mixture of pyrrole monomer (50 mg) and GO suspension (10 ml) was ultrasonically dispersed for 30 min, and then maintained at 60 °C for 24 h in an oven to facilitate the cross-linking of pyrrole with GO, resulting in a black cylindrical reduced graphene oxide hydrogel (RGOH). This RGOH was then rinsed with a 5:1 water/ethanol solution (by volume) for 24 h to remove residual pyrrole and polypyrrole oligomer. The reduced graphene oxide aerogel (RGOA) was subsequently obtained by freeze-drying of the RGOH for 24 h.

### 2.2 Fabrication of 3D porous graphene/ silicon oxycarbide monoliths

RGOA with attached nitrogen bearing pyrrole groups was further thermally reduced at 900°C for 1 h in a tube furnace under flowing argon at a heating rate of 180°C/h to combust the organic content and form N-doped reduced graphene oxide aerogel (NRGOA). The nitrogen doping here is conducted in order to reduce the systems oxygen content and further enhance electronic conductivity in the materials[53]. A solution of commercial preceramic polymer precursor, methylphenylvinylhydrogen polysiloxane (SILRES H62C, Wacker Chemie, Burghausen, Germany) was infiltrated into as-prepared NRGOA frameworks under low vacuum (200 mbar) to fabricate 3D graphene/polymer-derived ceramic architectures with varied PDC loadings. Towards this end, the polymeric precursor was combined with tert-Butanol (TBA, $(CH_3)_3OH$, Merck, Germany) at concentrations of 15, 25, 50, 100, and 300 mg/ml. Then NRGOA, which exhibits good wettability, was immersed into this solution under vacuum for 2 h to facilitate precursor penetration into the aerogel and eliminate bubbles. Subsequently, the wetted samples were taken out of the solution and freeze-dried for 12 h to remove the TBA. The resulting polymeric precursor loaded NRGOA (NRGOA-P) maintained their structure without visible changes as shown in Figure. 1, which was attributed to the use of high melting-point-solvent assisted infiltration and freeze drying instead of a more conventional solvent followed by vacuum drying. Subsequently, NRGOA-P cylinders were cross-linked at 250 °C for 2 h and pyrolyzed at 1000 °C for 3 h in a tube furnace under flowing argon at a heating rate of 100°C/h to obtain the final 3D N-doped graphene aerogel supported SiOC (NGA-SiOC) materials. Samples are denoted here as NGA-SiOCX, where X=15, 25, 50, 100, 300, corresponding to the values of precursor concentration. For comparison, unsupported SiOC particles were prepared by the pyrolysis of preceramic polymer H62C at the same heat-treatment conditions.

### 2.3 Characterization

X-ray diffraction (XRD) patterns were acquired using a Bruker D8 diffractometer (Bruker, Billerica, MA, USA) with Co Kα radiation (λ=0.1789 nm) over a 2θ range from 10° to 90°. Fourier transform infrared spectroscopy (FTIR) in Attenuated Total Reflection (ATR) mode was carried out in an Equinox 55 (Bruker, Germany) in the range of 500 – 4000 cm-1 to characterize bonding. X-ray photoelectron spectroscopy (K-Alpha TM, Thermo Scientific) was used to examine surface elemental composition. X-ray imaging (radiography) was performed using a microfocus X-ray source with a flat panel detector (Hamamatsu, Japan) with an area of ~ 120 x 120 mm² and a pixel size of 50 μm. Microstructural analysis was conducted by Scanning Electron Microscopy (SEM) using a Leo Gemini 1530 (Zeiss, Germany). Transmission electron microscopy (TEM) was used to confirm the presence of sandwiched multilayer structure of oxycarbide on graphene within the monolithic nanocomposite 3D NGA-SiOC.

### 2.4 Electrical characterization

Electrical resistances of NGA and NGA-SiOC composites were measured by a two-probe method using a digital multimeter (287, Fluke, US) with a sampling frequency of 2 Hz. The electrodes were painted with highly conductive silver paste (~ 0.01 Ω·cm) on both sides, to minimize contact resistance. The voltage-current (VI) curves were measured by a Probe Station II using Kiethley SCS4200 Semiconductor Parameter Analyzer (SCS4200, Kiethley, US).

### 2.5 Electrochemical measurements

The electrochemical characterization of electrodes based on the NGA-SiOC composites were tested using CR2032 stainless steel coin cells with lithium foils as the counter electrode. The active materials (85 wt%) were mixed with PVDF (10 wt%) and carbon black (5 wt%) in N-methyl-2-pyrrolidone to form a slurry. The working electrodes were prepared by coating the slurry on the surface of copper foil current collectors and drying at 100°C in a vacuum oven for 10 h to remove N-methyl-2-pyrrolidone. $LiPF_6$ (0.5 M) in ethylene carbonate and dimethyl carbonate (EC/DMC = 1:1) was used as the electrolyte. The cells were assembled in an argon-filled glovebox. The CV and GCD curves were measured using an electrochemical workstation (USA PARSTAT 273A) and a battery test system (LAND CT2001A) with a potential range between 0.005 and 2.5 V. Electrochemical impedance spectroscopy (EIS) measurements were done (as assembled) at 50% state of charging (SOC) with a Zahner ZENNIUM Impedance measurement unit under potentiostatic mode with 5mV amplitude between 100 mHz and 1 MHz. First, both cells were activated at 37 mA/g for five cycles. Then they were charged up to 50%SOC and rested there for 2 h before EIS measurement.

**3. Results and Discussion**

Highly porous NRGOA skeletons were produced via the polymerization-reduction process used here following freeze drying and thermal reduction. Figure S1 in the supplementary information illustrates the output of the methods used here. Reduced graphene oxide aerogel (RGOA) was synthesized via polymerization-reduction between pyrrole (Py) and graphene oxide (GO). The polypyrrole (PPy) grows preferentially on the surface of GO sheets due to electrostatic interactions between positively charged Py and negatively charged GO surfaces, π-π interactions between Py rings and conjugated segments in GO and hydrogen-bonding interactions[54]. Through these interactions, GO acts as a structuring agent with PPy forming a thin surface layer. The incorporation of PPy also changes the stacking of GO sheets from a 2D closely stacked layer to a 3D porous structure (Figure 2a, d), which was attributed to the increased ratio of attraction/repulsion forces that stabilizes the GO hydrogel network[55]. During further thermal reduction of RGOA, PPy decomposes, yielding ultralight NRGOA with extremely high compressibility, similar to that of RGOA (Figure S2). As shown in Figure 2e, the formed NRGOA exhibits a 3D structure with interconnected pores of dimensions ranging from tens to hundreds of micrometers, which provides sufficient space for the integration of the preceramic precursor into the NRGOA skeleton (Figure 2b). Monolithic NGA-SiOC composites were then formed through precursor infiltration, freeze-drying and high temperature pyrolysis (Figure 1). An infiltration technique referred to as *polymer–solvent solidification-sublimation* is used in place of conventional polymer–solvent evaporation in this work. During infiltration, the precursor attaches to NRGOA surfaces due to favourable wetting and adhesion. By using *tert*-butanol (TBA, melting point: 25ºC) as the solvent instead of other organic solvents (e.g. ethanol or tetrahydrofuran), freeze drying can be conducted following the infiltration process, which overcomes the problem of surface tension at the gas-liquid-pore wall. As a result, the 3D graphene skeleton maintains its morphology without shrinkage, collapse or agglomeration during infiltration and drying (Figure S3). After further pyrolysis, the scaffolds present uniform shrinkage and 3D graphene embedded polymer derived ceramic structures are obtained (Figure 2c).

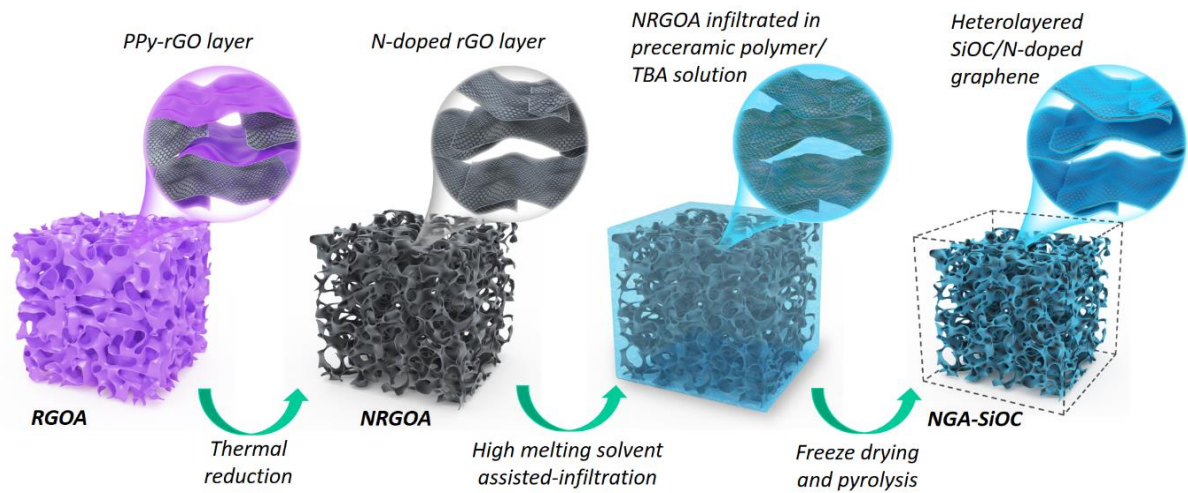

**Figure 1.** Schematic illustration of the fabrication of 3D NGA-SiOC porous monoliths.

The composition, geometry and microstructures of 3D porous NGA-SiOC monoliths can be tailored by adjusting the concentration of the H62C precursor in the TBA solvent. NGA-SiOC composites were fabricated here with NGA contents of 29.9, 19.8, 11.8, 6.4 and 2.5 wt%, corresponding respectively to precursor concentrations of 15, 25, 50, 100 and 300 mg/ml (Figure S4). The volume of 3D NGA-SiOC monolith gradually decreases with increasing precursor concentration (Figure S5). As shown in Figure S6, the NGA-SiOC monoliths present porous structures, with the number and size of pores inside gradually decreasing with increasing SiOC content. In the fabricated NGA-SiOC25 material (Figure 2f), interconnected graphene sheets are coated by thin SiOC layers, forming a heterolayered graphene/SiOC structure. The distribution of chemical elements is further studied by EDS mapping, as shown in Figures 2g-i, as well as S7, the uniform distribution of the SiOC throughout the NGA framework is confirmed by the matching elemental distribution patterns of oxygen and carbon. The oxygen content observed is present mainly in the pyrolyzed SiOC, as only sparsely distributed oxygen functional groups remain on NRGOA frameworks after thermal reduction. With increasing precursor concentration, pores space is filled to a greater extent by SiOC, entailing a decrease in the overall porosity (Figure S6).

In order to further investigate the distribution of SiOC on graphene at multiple length scales, NGA-SiOC25 samples were studied by radiography, SEM and HRTEM. Structural features are shown across three length-scales in Figures 2j-l. In x-ray radiographs, the RNGOA material shows little contrast, due to its composition of largely x-ray transparent carbon (Figure S8). Materials infiltrated with SiOC, revealed homogeneous X-ray attenuation as shown in the radiographs of Figure 2j and S8, which indicates that the SiOC layer on graphene surfaces is uniformly distributed throughout the interconnected graphene framework at the macroscale. NGA-SiOC300, in which pores are nearly entirely filled with SiOC, presents as solidly black in X-ray radiography. The micrograph in Figure 2k, reveals an individual structural unit of the NGA-SiOC composite scaffold, showing conformal growth of SiOC along the "wrinkled" morphologies of the graphene skeleton, confirming good compatibility between the two constituents of the composite material. The presence of a heterolayered structure of graphene and SiOC can be seen in Figure 2l. HRTEM further verifies the seamlessly coupled interface between the multilayer graphene and the amorphous SiOC of the heterolayered structure.

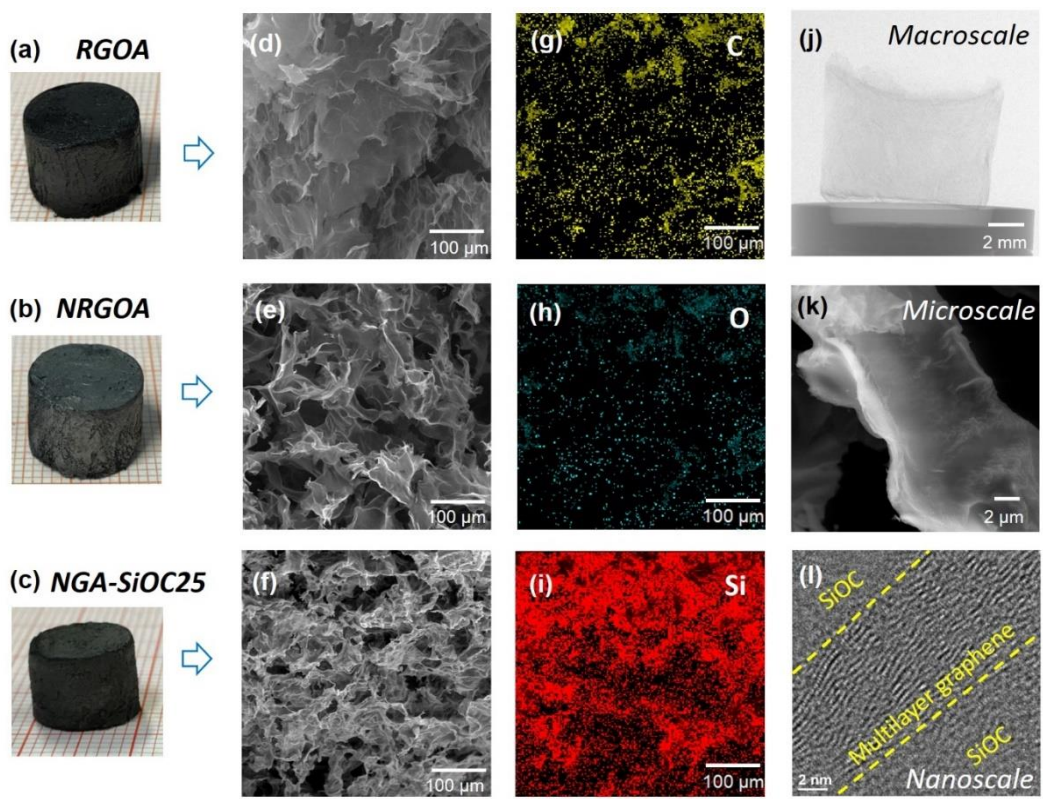

**Figure 2.** (a-c) Optical and (d-f) SEM images of RGOA, NRGOA and NGA-SiOC25, (g-i) EDS mapping of C, O, Si elements of (f), (j) radiography image of bulk NGA-SiOC25 (k) SEM image of an individual structural unit in the scaffold (l) high resolution cross-sectional TEM image revealing the graphene/SiOC sandwich structure of individual structural units in NGA-SiOC25.

Phase evolution in NGA-SiOC nanocomposites was characterized by XRD. The (001) crystallographic plane of GO is clearly evident in the diffractograms shown in Figure 3a, exhibiting an interlayer spacing of 8.33 Å, which is larger than that of graphitic carbon (which shows a major (002) reflection at 26.53°, corresponding to a spacing of 3.36 Å) owing to the oxygen present in GO and interlayer water molecules. After annealing at 900 °C for 1 h, the NRGOA showed a broad reflection at 2θ=26° corresponding to the (002) plane of graphitic carbon, suggesting the reduction of GO. As expected, no XRD reflections are observed in the diffractogram of SiOC, confirming the amorphous nature of polymer derived SiOC ceramics. The XRD patterns from NGA-SiOC15 and NGA-SiOC25 correspond to a combination of reduced graphene structure (i.e. graphitic carbon) and an amorphous phase. The disappearance of the graphitic reflection at 26.5° in pyrolyzed NGA-SiOC50 composites can be explained by the dominant amorphous SiOC constituent. However, the occurrence of XRD peaks at 26.5° from pyrolyzed NGA-SiOC300 composites is attributed to the formation of graphitic carbon domains, suggesting the decrease of the RGO sheet spacing through the removal of functional groups during pyrolysis.

Figure 3b presents ATR-FTIR spectra for GO, RGOA, and NRGOA. For the spectrum of GO, the bands at 1732, 1619, 1216, and 1040 cm$^{-1}$ correspond respectively to vibrations of carbonyl C=O stretching, C=C stretching, C–OH stretching, and epoxy C–O stretching. After polymerization-reduction, bands related to C=O, C–OH and C–O groups disappear almost completely, indicating the effective reduction of GO. For RGOA, the strong band around 1563 cm$^{-1}$ occurs on account of typical ring vibrations of PPy, and that at 1211 cm$^{-1}$ is due to the C−N stretching. The band at 1042 cm$^{-1}$ corresponds to in plane vibrations of C−H bonds, further confirming the existence of PPy in RGOA through interactions hydrogen bonding and π−π stacking between PPy and graphene[56]. ATR-FTIR analysis also confirms transformation of the H62C polymeric precursor, with its -CH$_3$, -CH$_2$=CH$_2$-, Si-C$_6$H$_5$ and Si-H groups, to SiOC ceramic through high temperature pyrolysis (Figure 3c, d)

The evolution of elemental composition and speciation of functional groups on specimen surfaces were examined using XPS (Figure S9). The intensity of the C 1s peak, located at 284.6 eV, continually increases from GO to RGOA and NRGOA, while the change of oxygen content (as indicated by the O 1s peak at ca. 532.4 eV) shows the opposite trend. GO exhibits a C/O atomic ratio of 1.9 with this value increasing to 5.44 through pyrrole mediated assembly, confirming the reduction of graphene oxide sheets. After further thermal reduction, the C/O atomic ratio of NGA reaches ca. 22.91, indicating that a majority of functional groups were eliminated. As summarized in Table 1, the surface elemental composition calculated from XPS was found to be Si= 18.6 at. %, O=34.57 at. % and C=46.83 at. % of pure SiOC, while the NGA-SiOC25 nanocomposite shows lower amounts of silicon (11.14 at. %), less oxygen (22.64 at. %) and a higher carbon content (64.03 at. %), with nitrogen present at 2.19 at. % after the introduction of NRGOA. Figure 4a shows the high-resolution C 1s spectrum of GO including C−C (284.6 eV), C−O (286.6 eV), C=O (287.8 eV), and O-C=O (289.1 eV), respectively. However, the peak intensity of C−O, C=O and O-C=O in RGOA after polymerization-reduction reaction decreases and a new carbon bond of C−N with the binding energy of 285.6 eV appears. After thermal removal of oxygenated functional groups in the form of gaseous species through the porous network, defect formation, lattice contraction, folding and unfolding of the layers and layer stacking, the oxygen groups further reduce and π-π shake up appears. In addition, a new bond of C-Si (283.7eV) emerges after encapsulation with SiOC, which dominates the main component in pure SiOC sample[57]. The Si 2p peaks were deconvoluted to five peaks of SiO$_4$ (103.6 eV), SiO$_3$C (102.8 eV), SiO$_2$C$_2$ (101.8 eV), SiOC$_3$ (100.8 eV) and SiC$_4$ (99.5eV) (Figure 4b)[58]. It is worth noting that the relative contents of SiC$_4$, SiOC$_3$ and SiO$_2$C$_2$ units decrease after the incorporation of NRGOA, especially with higher NRGOA content, which can be attributed to the fact that Si bonds more readily with O atoms than C atoms at high temperatures and the oxygen content was elevated by the remaining oxygen group from NRGOA (Figure 4c). A pronounced N 1s peak is observed for RGOA and NRGOA apart from the peaks for C and O elements. The high-resolution N 1s spectrum of the RGOA (Figure S10) reveals the presence of pyrrolic N, which further confirms the successful incorporation of PPy into the RGOA. After thermal reduction, N1s spectra in NGA can be deconvoluted into three peaks located at 398.2, 400.9, and 402.8 eV, which are attributed to pyridinic N, pyrrolic N, and graphitic N, respectively, suggesting N atoms are incorporated into the graphene network.

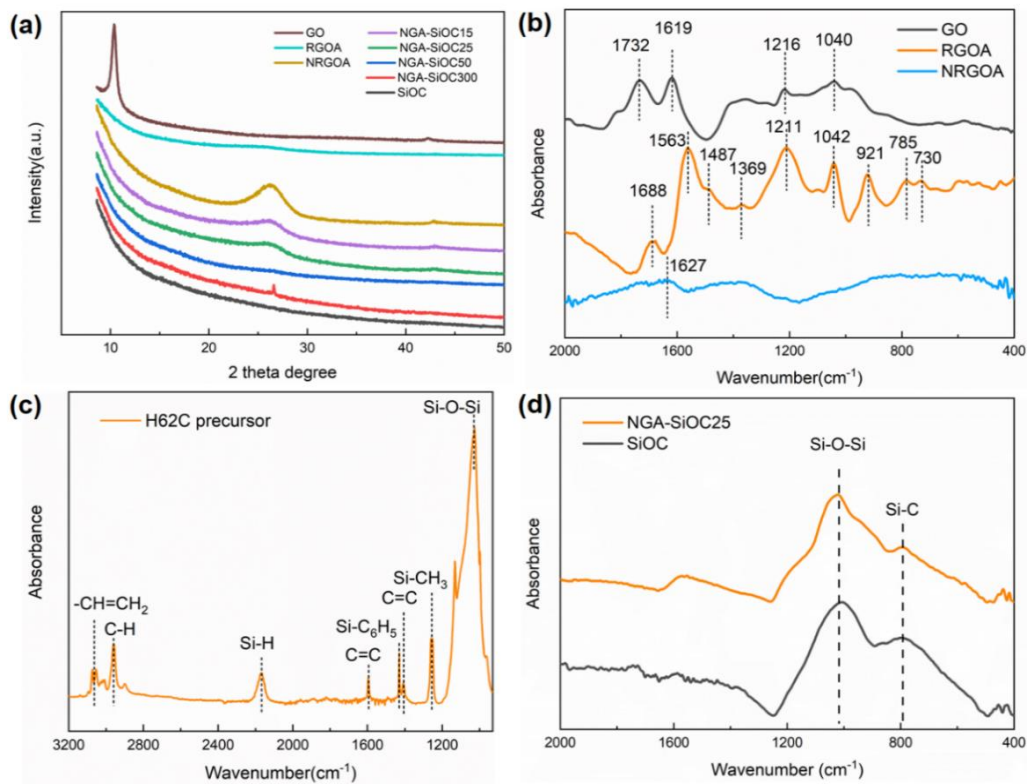

**Figure 3.** (a) XRD patterns of GO, RGOA, NRGOA, SiOC and NGA-SiOC nanocomposites. ATR-FTIR spectra of(b) GO, RGOA and NRGOA, (c) preceramic polymer and (d) SiOC and NGA-SiOC25

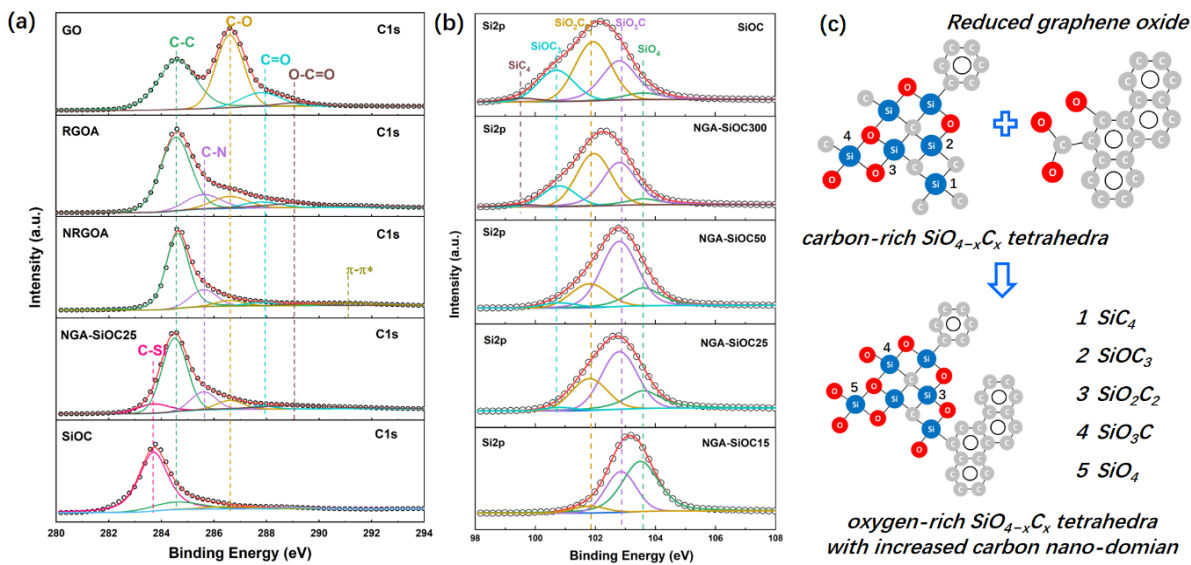

**Figure 4**. High resolution XPS spectra for (a) C1s of GO, RGOA, NRGOA, NGA-SiOC25 and SiOC, (b) Si2p of SiOC and NGA-SiOC nanocomposites, (c) Transformation of nanodomain structure and silicon coordination in SiOC after the introduction of RGOA.

**Table 1.** Atomic concentration of each constituent from the survey scans of GO, RGOA, NRGOA, NGA-SiOC nanocomposites and SiOC

| Sample | Si (atom %) | O (atom %) | C (atom %) | N (atom %) |
|---|---|---|---|---|
| GO | - | 34.39 | 65.61 | - |
| RGOA | - | 14.33 | 77.94 | 7.73 |
| NRGOA | - | 4.03 | 92.31 | 3.66 |
| NGA-SiOC15 | 11.03 | 24.85 | 61.69 | 2.44 |
| NGA-SiOC25 | 11.14 | 22.64 | 64.03 | 2.19 |
| NGA-SiOC50 | 12.83 | 27.27 | 58.31 | 1.59 |
| NGA-SiOC300 | 15.08 | 32.35 | 51.48 | 1.09 |
| SiOC | 18.6 | 34.57 | 46.83 | - |

Figures 5a-d show the porous microstructures and electrical conductivity of monolithic NGA-SiOC materials, demonstrating conductivities as high as 2.7, 8.4, and 85.5 S/m for NGA-SiOC15, NGA-SiOC50 and NGA-SiOC100, respectively, which are several orders of magnitude higher than those of pristine SiOC following pyrolysis at the same temperatures[59] and are of a similar magnitude to the conductivity of pure graphene aerogels[60-61]. Compared with other reported graphene/PDC composites summarised in Table S1 (SiOC-graphene@1500/1700 [50], SiOC-GO@1700 [47], RGO/CNTS/SiCN@1000 [26], GO/PSZ@800/1000 [51], SiCN-rGO@1000 [46], SiCNO-GO@1000 [48], SiCO/GO@1100 [62], rGO/Si$_3$N$_4$@1600 [63], SiCN/rGOA@1000 [49]), the lightweight, porous graphene/SiOC composites here present outstanding electrical properties at equivalent levels. The electronic conduction in NGA-SiOC is schematically illustrated in Figure 5e. The sandwiched multi-layer graphene sheets of the NGA scaffold provide well-interconnected pathways for electron transport at a large mobility. In contrast, graphene/PDC composites in which graphene is dispersed in the form of isolated particles in the ceramic phase lack such interconnectivity, and thus rely on free carbon in the ceramic matrix to facilitate electron percolation. In the aerogel infiltration approach applied here the NGA framework is retained intact, being free from damage due to its high mechanical robustness during the infiltration process, which effectively reduces possible deterioration (e.g., bubbles, weak interfaces, and cracks) in the microstructure. Each graphene sheet in the NGA provides a pathway for electron migration and contributes simultaneously to mechanical reinforcement. In this respect such conductive NGA-SiOC nanocomposites possess promising functionalities for electromagnetic shielding, as well as for application as thermoelectric / conductive ceramic composites.

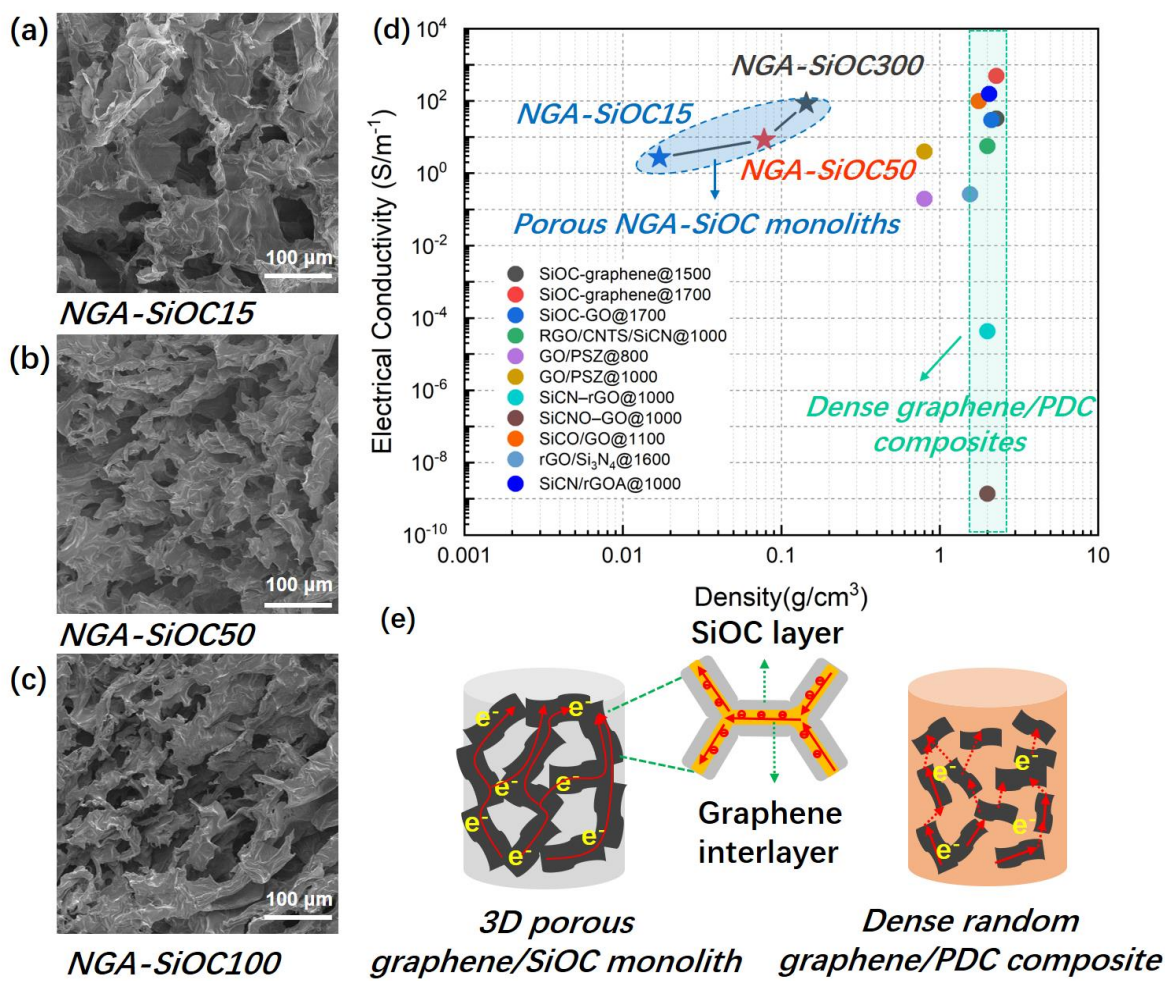

**Figure 5.** (a-c) SEM images of NGA-SiOC15, NGA-SiOC50 and NGA-SiOC100, respectively. (d) The electrical properties of the 3D porous NGA-SiOC monoliths (NGA-SiOC15, NGA-SiOC50 and NGA-SiOC100) and the reported graphene/PDCs (SiOC-graphene@1500/1700[50], SiOC-GO@1700[47], RGO/CNTS/SiCN@1000 [26], GO/PSZ@800/1000 [51], SiCN-rGO@1000[46], SiCNO-GO@1000[48], SiCO/GO@1100[62], rGO/Si$_3$N$_4$@1600[63], SiCN/rGOA@1000[49]) (e) Schematic of the conductive pathway for electron transport in porous graphene/SiOC monoliths and dense random graphene/SiOC composite. (Note: The unreported density of dense composites is defined as 2g/cm$^3$ in (d)).

The electrochemical performance of heterolayered graphene/SiOC was evaluated in half cells. Figure 6a shows the initial charge/discharge curves of NGA-SiOC nanocomposites of varied NGA content. Among them, the SiOC-NGA25 (heterolayered structure) delivers a high initial discharge capacity of 1116 mAh/g and a charge capacity of 751 mAh/g, with the highest initial coulombic efficiency of 67.3% (Table 2). Bare SiOC anodes (particles illustrated in Figure S11) show a similar discharge capacity of 1090 mAh/g, however the following charge only delivers 540 mAh/g translating to a coulombic efficiency of 49.5 %. The irreversible first-cycle capacity results from electrochemical formation processes such as solid-electrolyte interphase (SEI) layer formation[64].

Shao, Gaofeng, et al. "Polymer-Derived SiOC Integrated with a Graphene Aerogel As a Highly Stable Li-Ion Battery Anode." *ACS Applied Materials & Interfaces* 12.41 (2020): 46045-46056.

**Table 2.** Overview of the NGA content, first cycle discharging and charging capacity and initial coulombic efficiency for bare SiOC and NGA-SiOC composites electrodes

| Sample | NGA content (Wt.%) | 1st $C_{dis}$ (mAh/g) | 1st $C_{ch}$ (mAh/g) | Initial coulombic efficiency (%) |
| --- | --- | --- | --- | --- |
| NGA-SiOC15 | 29.9 | 796 | 479 | 60.2 |
| NGA-SiOC25 | 19.8 | 1116 | 751 | 67.3 |
| NGA-SiOC50 | 11.8 | 807 | 537 | 66.5 |
| NGA-SiOC300 | 2.5 | 878 | 539 | 61.3 |
| SiOC | 0 | 1090 | 540 | 49.5 |

The discharge potential is the highest for SiOC-NGA25 and decreases with decreasing NGA content with the lowest value for SiOC, indicating that the overpotential is lowered due to the NGA matrix. EIS measurements of NGA-SiOC25 and SiOC electrodes (Figure S12) give further insight into the single polarization processes. The ohmic part (summarized resistance of current collectors, electrolyte and electrodes, SEI) is similar for both samples showing values of 1.5 and 2.5 Ohm, mainly due to the electrolyte resistance. The IR-drop for a current at 37 mA/g corresponds to 0.6mV, which is negligible with respect to the reaction and diffusion polarization. The reaction polarization (charge transfer at the electrode/electrolyte interface), described by a semicircle in the Nyquist-plot, is around 30 Ohm for NGA-SiOC25 and ca. 150 Ohm for SiOC. According to the Butler-Volmer-equation the charge transfer reaction is related to the active surface area and exchange current density. Hence, the lower measured transfer resistance for NGA-SiOC25 indicates that the embedded multilayer graphene network imparts a higher exchange current density and improved electron transport. This directly influences the cell potential. This is emphasized by the reduced relaxation time constant of $\tau=75$ μs for SiOC25 compared to $\tau=734$ μs for SiOC. This relaxation time constant ($\tau=RC$) describes the time it takes a process to return to equilibrium in the absence of the electric field. Smaller $\tau$ values correspond to faster processes. The diffusion polarization is related to the sloping line in the low frequency range of the Nyquist-plot and describes the diffusion of Li$^+$ ions in electrode's vicinity. A rough estimation of the Warburg diffusion W gives 50 Ohms$^{-0.5}$ and 150 Ohms$^{-0.5}$ for NGA-SiOC25 and SiOC, respectively. Lower W corresponds to a lower diffusion polarization, due to a higher diffusion coefficient, surface area and/or ion concentration. The lower reaction and diffusion polarizations explain the higher discharge potential and lower charge potential for NGA-SiOC25 compared to SiOC. The higher dis-/charging overpotentials of the pristine SiOC anodes lead to a reduced capacity, with the designated cut-off potentials of 5mV and 2.5 V (for discharging and charging, respectively) being reached earlier.

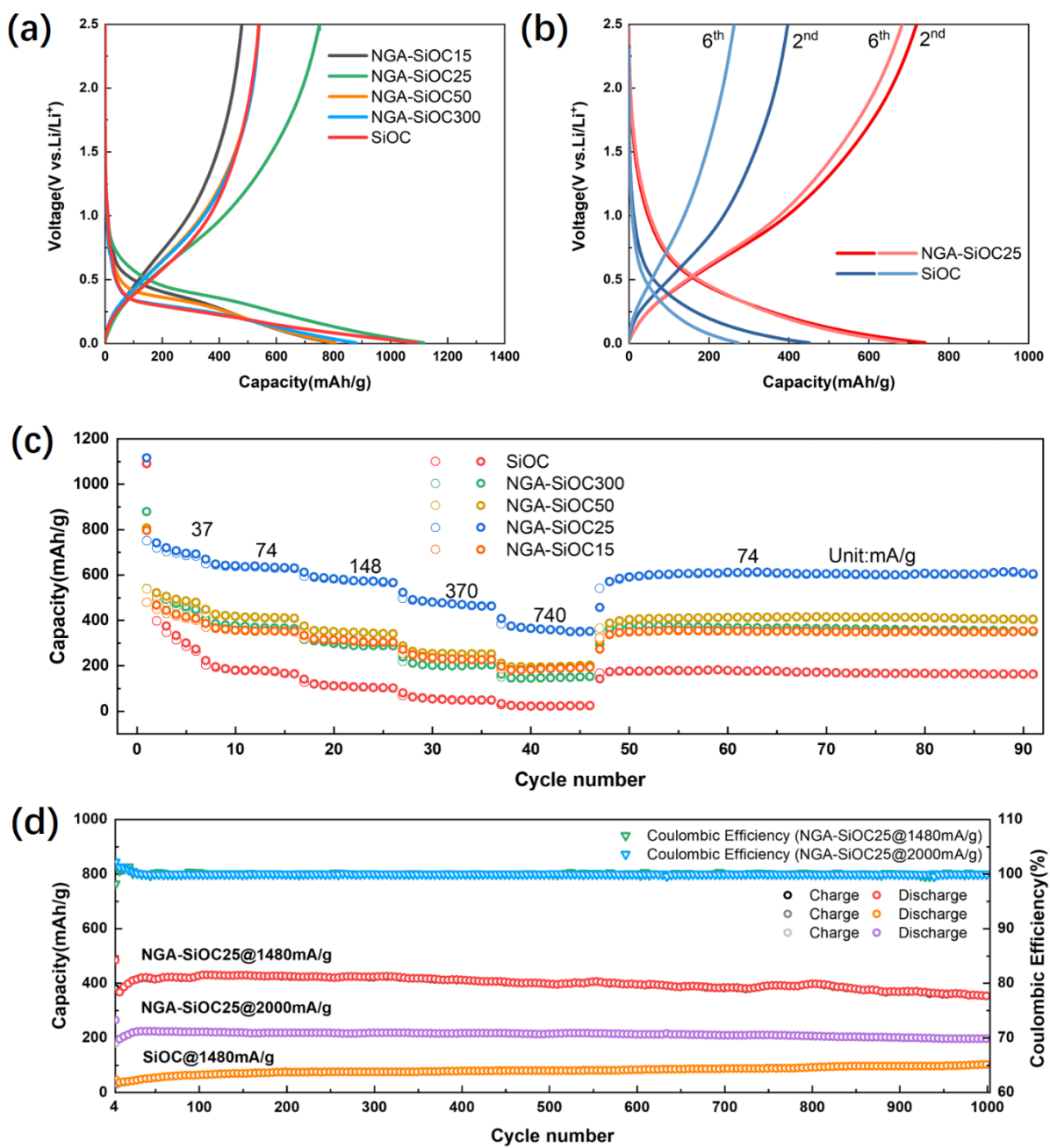

**Figure 6.** Galvanostatic charge–discharge profiles in the 0.005–2.5 V window (versus Li/Li$^+$) for (a) 1st cycle of NGA-SiOC nanocomposites and SiOC at 37mA/g (b) 2nd to 6th cycles of NGA-SiOC25 and SiOC at 37mA/g. (c) Rate performances of NGA-SiOC nanocomposites at various current density, (d) Cycling performances of NGA-SiOC25 at current density of 1480 and 2000mA/g, and SiOC at current density of 1480 mA/g. (37 mA/g for the first 3 cycles)

The difference between the cycling stability of SiOC and NGA-SiOC25 anodes in the first 6 cycles is shown in Figure 6b. Compared to the SiOC anode, the NGA-SiOC25 anode demonstrates much better cycling stability and stable capacity at 37 mA/g from the 2$^{nd}$ to 6$^{th}$ cycle. Although both electrode types exhibit a significant decrease in capacity over the first six cycles, the capacity drop of SiOC is more significant, possibly because of its greater propensity to crack as the result of weak bonding between the PVDF binder and the active phase.[23] In composite materials, the graphene phase imparts mechanical robustness and may reduce the formation of performance degrading cracks in the SiOC component that occur during cyclical structural changes associated with lithiation and delithiation of the anode.

A further explanation for the drop in capacity is the varying tendency of certain $SiOC_x$ phases to undergo irreversible lithiation. It has been reported that the dominant oxygen-rich phases like $SiO_2C_2$, $SiO_3C$, and $SiO_4$ undergo a reversible reaction with Li-ions, while carbon-rich phases of $SiOC_3$ and $SiC_4$ result in irreversible reactions with lithium[65]. Here, XPS results show that oxygen-rich phases are more dominant in NGA-SiOC25 than in SiOC. And thus the stronger decay of capacity in SiOC within the first cycles could be attributed to the higher extent of irreversible lithiation reactions with carbon-rich phases.

Figure 6c presents the galvanostatic cycling performance of the NGA-SiOC nanocomposites and SiOC at different charge/discharge current densities, illustrating the current rate capabilities and cycling performances for materials of different NGA contents. NGA-SiOC25 shows a superior rate capability among the investigated NGA-SiOC anode materials. It exhibits discharge capacities of 628, 565, 462 and 351mAh/g in the 10th-cycle at 74, 148, 370 and 740 mA/g, which are much higher than the SiOC anode (164, 102, 49 and 24 mAh/g, respectively). The discharge capacity of NGA-SiOC25 is significantly enhanced compared to that of SiOC and other NGA-SiOC nanocomposites at the same current density. More importantly, for NGA-SiOC25, when the current density is reset to 74 mA/g after cycling at different current densities, the discharge capacity is restored to 605 mAh/g and remains stable during the following cycles at 74 mA/g.

With regards to the phase content of the NGA-SiOC composites, it was found that either increasing or decreasing the amount of SiOC (by varying the concentration of the preceramic polymer precursor) relative to that of the NGA-SiOC25 sample, results in a decrease of electrochemical performance, in terms of initial coulombic efficiency, reversible capacity and current rate capability. This can be understood as insufficient SiOC content correlates to a lower fraction of PDC phase and thus a reduced specific capacity. Excessive SiOC infiltration results in lower porosity, poorer reaction kinetics and increased overpotentials, as shown in Figure 6a. Furthermore, the presence of thick layers of SiOC on graphene surfaces brings about a higher level of oxygen-lean phases and thus, as mentioned earlier, greater levels of irreversible lithiation leading to low reversible capacity. Thus, it is important in these materials to control the ratio of NGA to SiOC to produce an optimum result. In spite of these factors, all NGA-SiOC nanocomposites investigated here display better electrochemical properties than pure bulk SiOC. The NGA-SiOC25 electrodes also exhibit the best capacitance retention characteristics among the nanocomposites studied, which further confirms the inherently stable microstructure of NGA-SiOC25 nanocomposite during the electrochemical reactions. In addition, the coulombic efficiency of NGA-SiOC25 is approximately 99−100% after the initial cycle in which the SEI layer is formed, indicating its superior capacitive reversibility for the repeated insertion/ deinsertion processes that are very critical for practical and manufacture-scaled applications.

To date, the cycling performance of SiOC based anodes at high-rates has not been significantly

Shao, Gaofeng, et al. "Polymer-Derived SiOC Integrated with a Graphene Aerogel As a Highly Stable Li-Ion Battery Anode." *ACS Applied Materials & Interfaces* 12.41 (2020): 46045-46056.

explored. In this work, the cycling performance of bulk SiOC and the NGA-SiOC25 nanocomposites was further examined with 1000 galvanostatic charge/discharge cycles at high current densities of 1480 and 2000 mA/g, as shown in Figure 6d and S13. The composite possesses a reversible capacity of 352 mAh/g with a coulombic efficiency of 99.97% after 1000 consecutive cycles. The capacity retention was ~95% of the second cycle capacity, which is significantly improved in comparison to that of pure and modified SiOC anodes reported in previous studies, as shown in Figure 7 and Table S2. As shown in Figure 8, there are no obvious cracks among layered graphene-SiOC structures and no segregation between graphene and SiOC after 1000 cycles at 1480mA/g, indicating a good cycling lifetime. Even at a current density of 2000 mA/g, a reversible capacity of 196 mAh/g is attained in the nanocomposite with a coulombic efficiency of 99.96% after 1000 consecutive cycles.

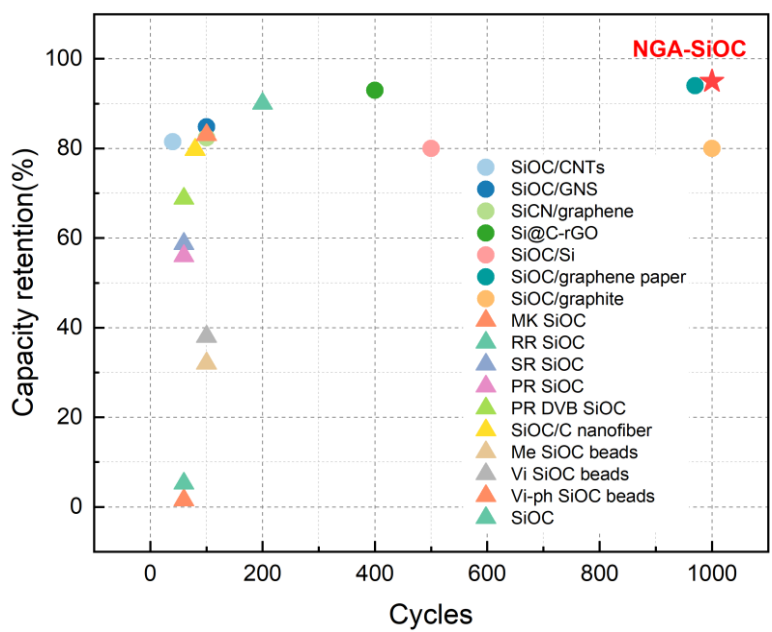

**Figure 7**. The cycling performace (capacity retention) of the NGA-SiOC25 anode from this work, compared with performance of similar materials reported in earlier studies. (Full data in supplementary material)

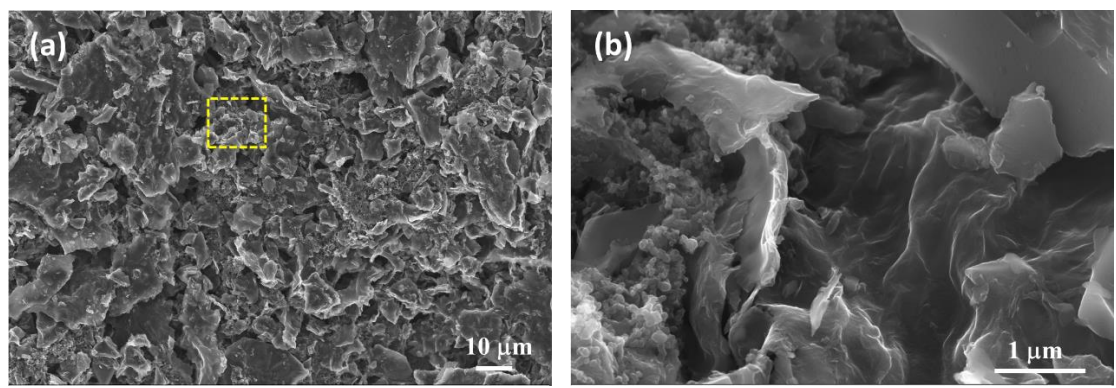

**Figure 8.** SEM images of NGA-SiOC25 anode after 1000 cycling at 1480mA/g

To examine the electrochemical performance of the NGA-SiOC nanocomposites fabricated here, the specific capacity values and stability of comparable SiOC based anode materials reported in the literature are listed in Table S3. It can be seen that the specific capacities attained in NGA-SiOC25 materials at 1480 and 2000mA/g are significantly higher than similar systems studied previously[18, 21, 42-44]. The favourable capacities and cycling behaviour, even at high rates, as achieved in the presently reported composite materials can be interpreted across different identifiable length-scales shown in Figure 9; (1) Molecular scale: the elemental composition of $SiO_{4-x}C_x$ tetrahedral units is tailored by oxygen groups of reduced graphene oxide from carbon-rich to oxygen-rich $SiO_{4-x}C_x$. As carbon rich $SiOC_3$ units react irreversibly with lithium to form $SiC_4$ units when lithium is inserted, the increasing oxygen content manifests in higher reversible capacities as shown here. (2) Nanoscale: the presence of increased levels of free carbon domains within the PDC provide higher levels of Li-ion intercalation sites and raise the intrinsic conductivity of the SiOC component in the composite. (3) Microscale: the graphene network in which the SiOC is embedded facilitates high electrical conductivity and ion mobility, shortens diffusion distances and enables the full utilisation of the SiOC material.

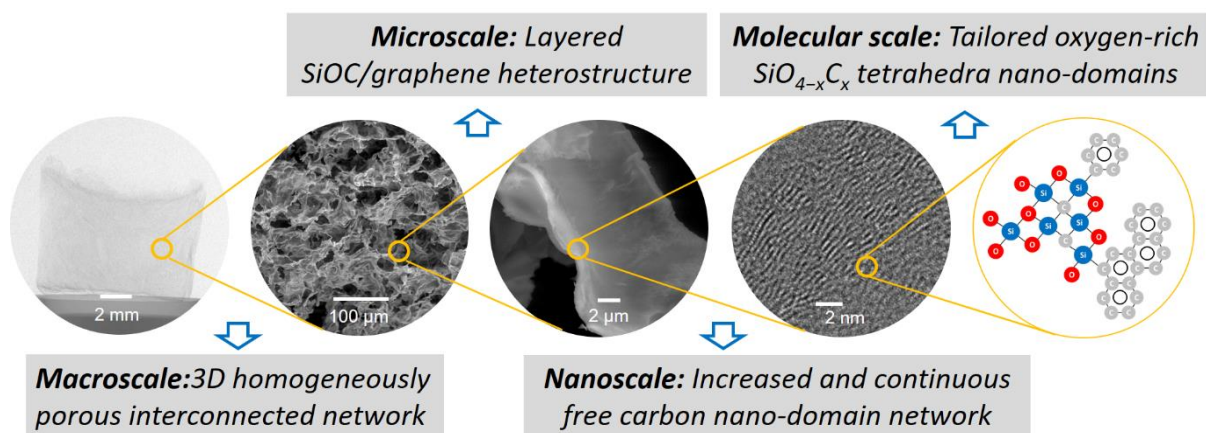

**Figure 9.** Schematic illustration of structure / performance relationships across multiple length-scales.

## 4. Conclusion

We have reported here a method for the infiltration-based fabrication of monolithic composites consisting of a polymer derived SiOC ceramic on surfaces of a 3D porous graphene structure. The highly porous and mechanically stable N-doped reduced graphene oxide aerogel, which was formed by polymerization-reduction and freeze drying, provides sufficient space for the integration of a preceramic precursor into its 3D skeleton. Facilitated using a relatively high-melting-point solvent, 3D porous NGA-SiOC monoliths were then fabricated via a polymer–solvent solidification-sublimation infiltration followed by high temperature pyrolysis. In 3D NGA-SiOC materials, thin SiOC layers are homogeneously coupled with multilayer graphene sheets, forming heterolayered graphene/SiOC structures. NGA-SiOC nanocomposites fabricated here demonstrate low density and high electrical conductivity, which is attributed to the NGA skeleton as well as the presence of increased levels of free carbon within the oxycarbide phase. It is postulated that the excellent cycling performance and high charging rates found here are also facilitated by these structures, which provide rapid and extensive reversible lithiation sites. The encouraging results found in this work motivates further investigations into graphene network based PDC composites and their optimisation with a view towards Li-ion battery materials.

**Acknowledgements**

This work was supported by China Scholarship Council (CSC, 201608320159, 201604910900), Deutsche Forschungsgemeinschaft (GU 992/17-1), The Startup Foundation for Introducing Talent of NUIST. We also thank Dr. Paul H. Kamm for conducting the X-ray imaging measurement, Dr. Franziska Schmidt for conducting the SEM measurement, and Dr. Sören Selve for conducting the TEM measurement, respectively.

**Supporting information:**

Supplementary graphics providing further information relating to the macrostructure, microstructure, composition and performance of materials fabricated in the present work

Supplementary tables providing a comprehensive comparison of the conductivity and cycling performance of anode materials studied here with related studies reported in the literature.

Shao, Gaofeng, et al. "Polymer-Derived SiOC Integrated with a Graphene Aerogel As a Highly Stable Li-Ion Battery Anode." *ACS Applied Materials & Interfaces* 12.41 (2020): 46045-46056.